# Software Development with Scrum: A Bibliometric Analysis and Profile

Peter Kokol
peter.kokol@um.si

Sašo Zagoranski
*Semantika*, saso.zagoranski@semantika.si

Marko Kokol
*Semantika*, marko.kokol@usemantika.si

# Software Development with Scrum: A Bibliometric Analysis and Profile

Peter Kokol, Marko Kokol, Sašo Zagoranski

*Faculty of Electrical Engineering and Computer Science, University of Maribor, 2000 Maribor, Slovenia, peter.kokol@um.si*

*Semantika, 2000 Maribor, Slovenia, marko.kokol@um.si*

**ABSTRACT**

*Introduction of the Scrum approach into software engineering has changed the way software is being developed. The Scrum approach emphasizes the active end-user involvement, embracing of change, and /iterative delivery of products. Our study showed that Scrum has different variants or is used in combination with different methods. Some tools not normally used in the conventional software approaches, like gamification, content analysis and grounded theory are also employed. However, Scrum like other software development approach focuses on improvement of software process, software quality, business value, performance, usability and efficiency and at the same time to reduce cost, risk and uncertainty. Contrary to some conventional approaches it also strives to boost soft factors like agility, trust, motivation, responsibility and transparency. The bibliometric synthetic scoping study revealed seven main research themes concerned with the Scrum research.*

**INTRODUCTION**

Introduction of the agile approach especially Scrum into software engineering has changed the way software is being developed. The Scrum approach emphasizes the active end-user involvement, embracing of change, and evolutionary/iterative delivery of products. Among agile methods Scrum seems to be employed in most countries and by most of software development companies [1]. Scrum is not only used in software development. It is implemented as a strategy teamwork and also as a teaching tool in educational institutions [2], and in agile maturity model research [3]. Historically Scrum research has its early roots in software engineering publications introducing innovative solutions to management, and development of software products [4]–[6]. Later historical roots were concerned with Scrum development process [7], software development management with Scrum [8] and empirical research of Agile software development [9].

Over the course of the conceptual evolution, the research on Scrum has attracted considerable attention in almost one and half thousand publications. The research literature has evolved in importance, quantity and complexity. However very few attempts have been made to holistically synthetize the research findings. The few published reviews were devoted to specific Scrum topics like the language used of Scrum communications [10], testing practices in software quality [11] and a bibliometrics study on analysing the research literature of Scrum impact on the productivity and efficiency of software development [12]. . Consequently, there is an extensive body of literature reporting on Scrum research, that has yet to be characterized. To close this gap we performed a study aimed to scope and chart the existing literature on the Scrum research, taking into account Scrum unique characteristics.

Scoping reviews have become increasingly popular as a form of knowledge synthesis [13]. They have great utility for synthesizing research evidence and are often used to categorize or group existing literature in a given field in terms of its nature, features, and volume. Scoping also takes into account the availability of resources. In the fast developing field like software engineering where the knowledge develops extremely rapidly and massively the knowledge synthesis reports have to be produced in a timely manner. To speed the scoping process even further and to enable the analysis of all research publications regarding Scrum, distant reading in combination with bibliometric mapping instead of manual reading of publications [14], [15] was used, thus making scoping process synthetic and semiautomatic. The objective of this synthetic scoping review was to identify and report on the extent literature concerning Scrum research, to map the Scrum research into the landscape of themes and second to identify which methods, tolls, roles, artefacts and objectives emerged in the scope of determined themes.

**RESEARCH METHODOLOGY**

Scoping reviews can be as a type of research synthesis that aims to 'map the literature on a particular topic or research area and provide an opportunity to identify key concepts; gaps in the research; and types and sources of evidence to inform practice, policymaking, and research' [16] or more simply as a process of mapping the existing literature or evidence [17]. Scoping and systematic reviews are similar in the manner that they both use rigorous and transparent methods to identify and analyse relevant literature to answer research questions [18]. However they also differ considerably in several ways. In scoping reviews the research questions are normally broader, inclusion/exclusion criteria can be developed post-hoc, data extraction is not so detailed and qualitative synthesis is done more frequently than quantitative [19]. Scoping reviews deal with broader topics, might include different types of studies, do not evaluate the quality of the studies, create narratives to sum the results, and consider resources limitations like cost and time [20]. Additionally, data extraction is focused on key general themes. Indeed, the knowledge synthesis is usually expressed as a literature overview from in which general themes are identified [21].

Scoping reviews are performed when there are research gaps in specific areas of research, to disseminate the research results, to clarify a complex concept, as a preliminary step before conducting systematic reviews to explore the extent of literature, and, finally, when informative summary of a research in a specific professional area is needed [17], [22]–[25]. Recently, scoping reviews are also employed for knowledge translation [26]. Arksey and O'Maley [17] proposed a six step methodological framework to perform scoping reviews. This framework was later extended by Levac, Colquhoun & O'Brien [23] and Joanna Briggs Institute [27].

In this study we used the scoping framework proposed by Arksey and O'Malley [17] and the suggestions encased by Levac et al. [23]. The framework was triangulated with bibliometric mapping and thematic analysis to make the scoping semi-automated i.e. synthetic [28], [29].In this manner the review proceeded trough following phases: identifying the research question, harvesting relevant publications, charting and mapping the publications metadata using the above triangulation approach, and collating, summarizing, and reporting the results. The optional 'consultation exercise' was not conducted.

**Identifying the Research Question**

This review was guided by the question: *What research themes emerge in the Scrum scientific literature and which methods, tolls, roles, artefacts and objectives appear in the context of the themes*.

**Harvesting Relevant Publications**

After experiment with various search stings the corpus was finally harvested from the Scopus database on 24[th] of February 2020, using the search string *Scrum AND agile AND software*. Our study was focused more on the basic research and not so much of the application of Scrum in other disciplines Tso we limited the search to the subject area *Computer Science*. No further limitations i.e. on date, language, or publication type were set. The advantage synthetic scoping is also to enable that no traditional eligibility criteria need to be set, ineligible terms or author keywords are excluded during the thematic analysis. Hence, they don't appear in charting and mapping activities and don't influence the results.

**Charting and Mapping the Publications Metadata**

The publications metadata was analysed first to determine the extent of the literature production in the manner to identify most prolific countries, institutions and source titles, and literature production dynamics. For the bibliometric mapping we used the VOSViewer software [30]. VOSviewer has been used to create bibliometric maps (also called science landscapes) in various studies.[31] The VOSviewer software has powerful visualization capabilities, meaning that resulting can be presented in various forms of bibliometric maps, consequently emphasising different aspects of the analysed research literature like term cluster landscapes, collaboration or citation networks, bibliometric coupling and similar. VOSviewer first uses text mining functions to identify relevant text elements than analyses them using an unified approach for both mapping and clustering. The approach is based on a normalized term co – occurrence matrix and a similarity measure which calculates association strength between terms [32]. In this manner, the VOSviewer software merges text elements that are closely associated into clusters. The proximity of the text element can be interpreted as an indication of their similarity. VOSviewer. The popularity of text element is indicated by the size of the element in the map.

Using a customized Thesaurus file, we excluded the common terms like study, significance, sample, baseline, group, experiment and eliminated geographical names and time stamps. The cluster landscapes were used to identify the themes of Scrum research employing thematic analysis [33]. The terms and keywords were used as codes during thematic analysis [34]. First, the term cluster landscape was analysed, and the clusters denoted with appropriate themes. Next the similar analysis was performed for the keyword cluster landscape, and the results of both analyses were collated. During the next step both landscapes were further analysed to determine which methods, tolls, roles, artefacts and objectives were used/targeted in each of the determined themes.

Finally, the most cited papers for the determined methods, tolls, roles, artefacts, and objectives per theme were identified using the citation data from Scopus. The results were collated and summarised in a tabular form. The hot topics were identified by an approach devised by Kokol at al [35].

## RESULTS

The search resulted in 1279 papers. Among them there were 949 conference papers, 217 original articles, 60 conference reviews, 33 book chapters, 11 reviews, five books and four other types of papers. The first two papers indexed in Scopus were publish in 2002 and were associated with integrating agile programming in software engineering courses [36] and the use of Extreme programming in web application development [37].

The dynamics of the overall literature production (Figure 1), shows that in the incubation period (2002 – 2005) the trend in the number of publications was positive, but slow. After 2005 the steeper began, lasting till 2016, when a small decrease in productivity could be observed. In 2017 the growth begun again reaching the peak in 2019 with 162 publications. The average growth was 11.1 publications per year. Regarding journals the more intensive growth in productivity began after 2010, but the trend was irregular. The peak was reached in 2019 with 26 articles. Most of the publications appeared in conference proceedings, which may imply that the Scrum research didn't yet establish a list of core journals to report research results. On the other hand in the fast development field like Scrum based software development, the conferences might be a more appropriate venue for exchange and maturation of ideas.

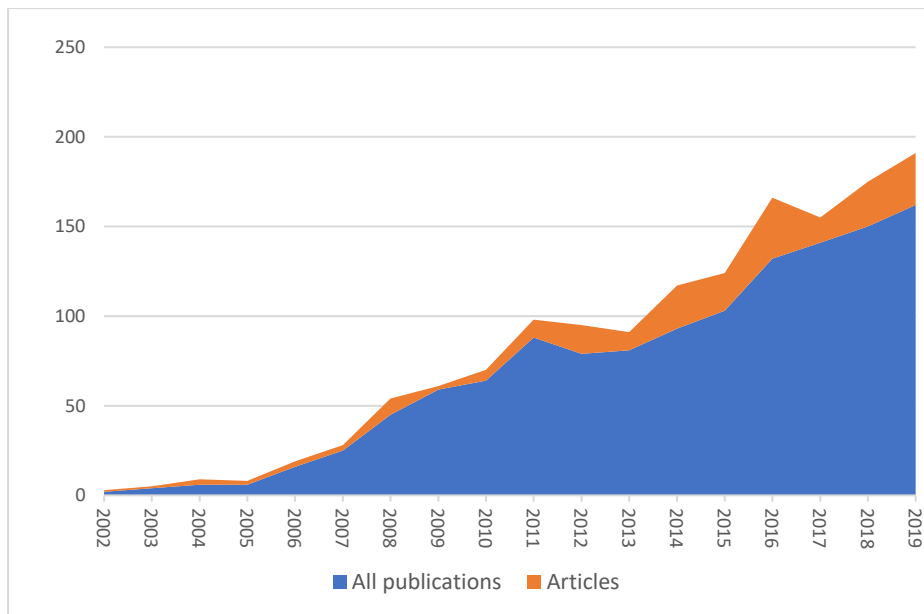

*Figure 1 The dynamics of research literature production*

**Extent of the Literature Production**

As stated above, much of the research on Scrum was presented on conferences and published in conference proceedings. The most productive conferences were Lecture Notes In Business Information Processing (n=98), Lecture Notes In Computer Science Including Subseries Lecture Notes In Artificial Intelligence and Lecture Notes In Bioinformatics (n=90), ACM International Conference Proceeding Series (n=72), Communications in Computer And Information Science (n=57), Advances In Intelligent Systems And Computing (n=42), Ceur Workshop Proceedings (n=30), Proceedings International Conference On Software Engineering (n=27), Procedia Computer Science (n=15), Proceedings Frontiers In Education Conference Fie (n=10) and Conference On Human Factors in Computing Systems Proceedings (n=5). Most of the above proceedings are from computers science or related fields. The publisher are internationally recognised entities and the proceedings are highly cited, mostly they have the H Index above 40. The highest H Index was reached by the Lecture Notes In Computer Science with H Index = 324.

The most prolific journals were IEEE Software (n=21), Information And Software Technology (n=20), Journal Of Systems And Software (n=19), Empirical Software Engineering (n=10), Journal Of Software Evolution And Process (n=10), Crosstalk (n=7), International Journal Of Advanced Computer Science And Applications (n=7), Journal Of Theoretical And Applied Information Technology (n=6) and International Journal of Engineering and Advanced Technology (n=5). The journals are mostly from the Computer science or Information Technology categories. In general their Scimago JCR factors (Scopus, Elsevier, Netherlands) are above 0.2, especially for the top productive journals and they are widely recognised as the top journals in software engineering. The less productive journals are not yet so recognized and have lower, however increasing JCR factors and might become core archival journals for the Scrum publications.

The most productive countries were United States (n=169), Brazil (n=124), Germany (n=108), India (n=76) and Finland (n=64), Sweden (n=51), Norway (n=47), United Kingdom (n=47), Netherlands (n=46) and Spain (n=43). United States is the country producing the most software globally [38]. USA employs the largest number of programmers (cca. four millions) globally and Germany the largest number of programmers in Europe (cca. 850000) [39]. Brazil and India are the best countries in which to outscore software development [40]. USA, Norway, Sweden, Germany and Finland and UK are among the countries where software developers have highest salaries [41]. Most of the top productive countries in Scrum research are also the most productive countries in the total scientific output i.e. USA is the first, UK third and Germany fifth.

The most productive organisations are Aalto University (n=23), SINTEF Foundation for Scientific and Industrial Research (n=20), Instituto Tecnologico de Aeronautica,(=17), University of Limerick (n=16), SINTEF Digital (n=16), Technical University of Munich (n=15), Lero - The Irish Software Engineering Research Centre (n=14), Fraunhofer Institute for Experimental Software Engineering IESE (n=13), Reykjavik University (n=13), Norges Teknisk-Naturvitenskapelige Universitet (n=12) and Universidade Federal de Pernambuco (n=12). Despite the fact that USA is the most productive country none of the most productive institutions is located there. The most productive institutions are situated in Europe and Brazil. It is interesting to note that three of the most productive institutions are located in Norway. These facts might imply that Scrum research in USA is spread through many institutions while in Europe and Brazil it is mostly done in very strong research centres.

**Thematic analysis**

Seven clusters emerged in the term cluster landscape (Figure 2), and 10 cluster in the author keywords cluster landscape (Figure 3). During the thematic analysis some of the clusters from the author keywords landscape were fused resulting in seven final themes compiled in the Table 1.Those seven theme are: Improving software processes with education and training; Interacting with users to develop working software, Other software development methods and Scrum, Translation of Scrum development approach, Interactive requirements engineering with user stories, Managing Scrum team in global software development and Scrum activities in every day work. Themes reveal that Scrum research is very wide. The research range from every day software development to the development in global and distributed environments. The research focus is frequently devoted to the topics where Scrum differs from traditional software development approaches. A lot of research is also dedicated to how to incorporate the Scrum into traditional software engineering courses and how to translate Scrum paradigm into practice. The study also revealed that Scrum research is frequently performed in the context of other software development approaches, either to compare Scrum with other approaches or to present frameworks how Scrum can be used in combination with other development approaches to improve software development process.

Quantitatively the most publications were concerned with the research on how first to interact with the users and stakeholders aiming to develop the best possible software system and second to translate the Scrum into practice. This research in general followed the directions set in the Agile manifesto [42]. Less research was done on how to introduce Scrum into the software engineering curricula and courses. The researchers mainly focused on how to teach and train Scrum users outside the traditional educational systems.

During the analysis of which methods, tolls, roles, artefacts and objectives were used/targeted in single themes (Table 2) we encountered that Scrum has many variants (or is used in combination with other methods), It is used in combination with other agile approaches, like Feature driven development [43], Xtreme programming [37] or Crystal Clear [44]. Additionally, it is combined with more traditional and systemic software development approaches like Dynamic System Development Method [45], Rational Unified Process [46] or Lean Development [47]. Some tools not normally used in conventional software approaches, like gamification, content analysis and grounded theory are also employed during software development. However, Scrum like other software development approach focuses on improvement of software process, software quality, business value, performance, usability and efficiency and at the same time to reduce cost, risk and uncertainty. Contrary to some conventional approaches it also strives to boost soft factors like agility, trust, motivation, responsibility and transparency.

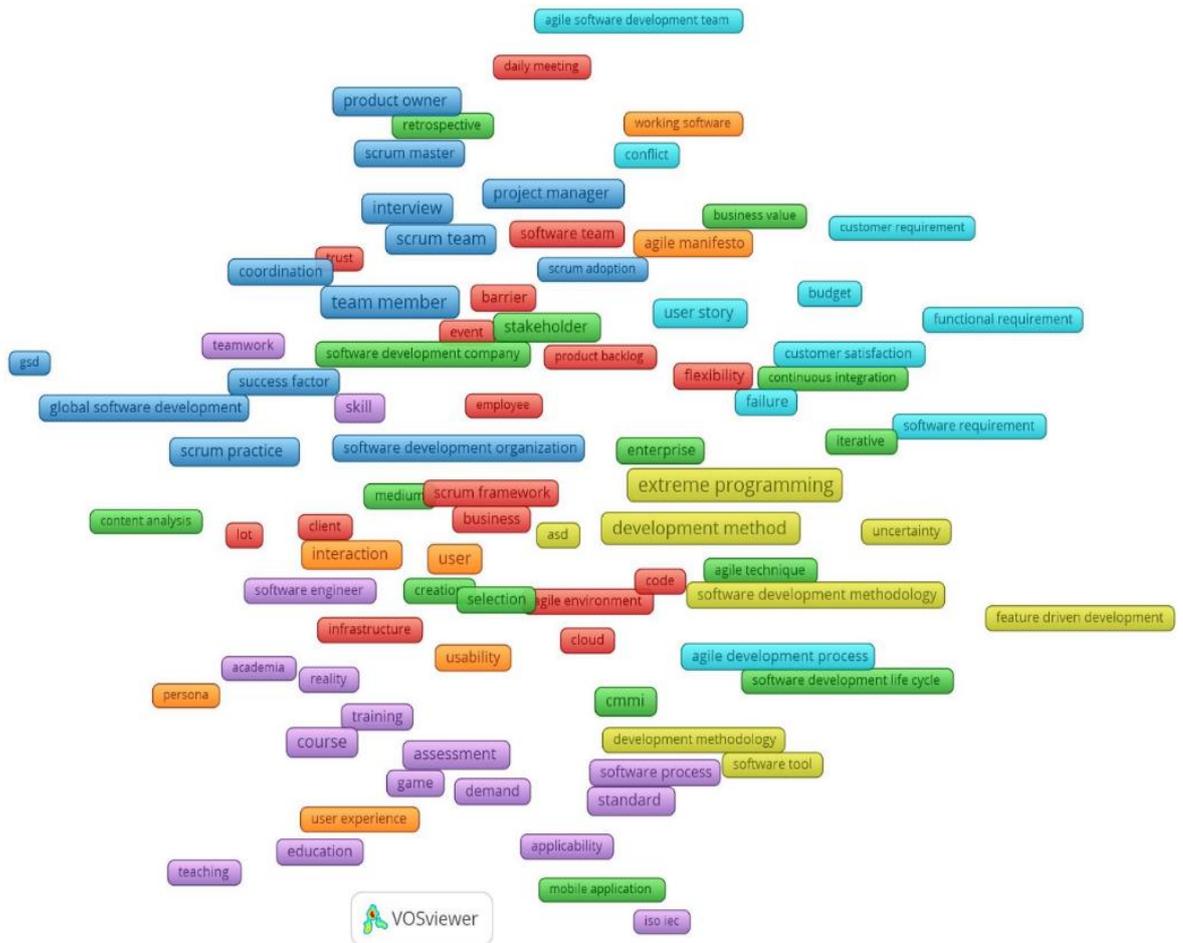

*Figure 2. The term cluster landscape of the Scrum research*

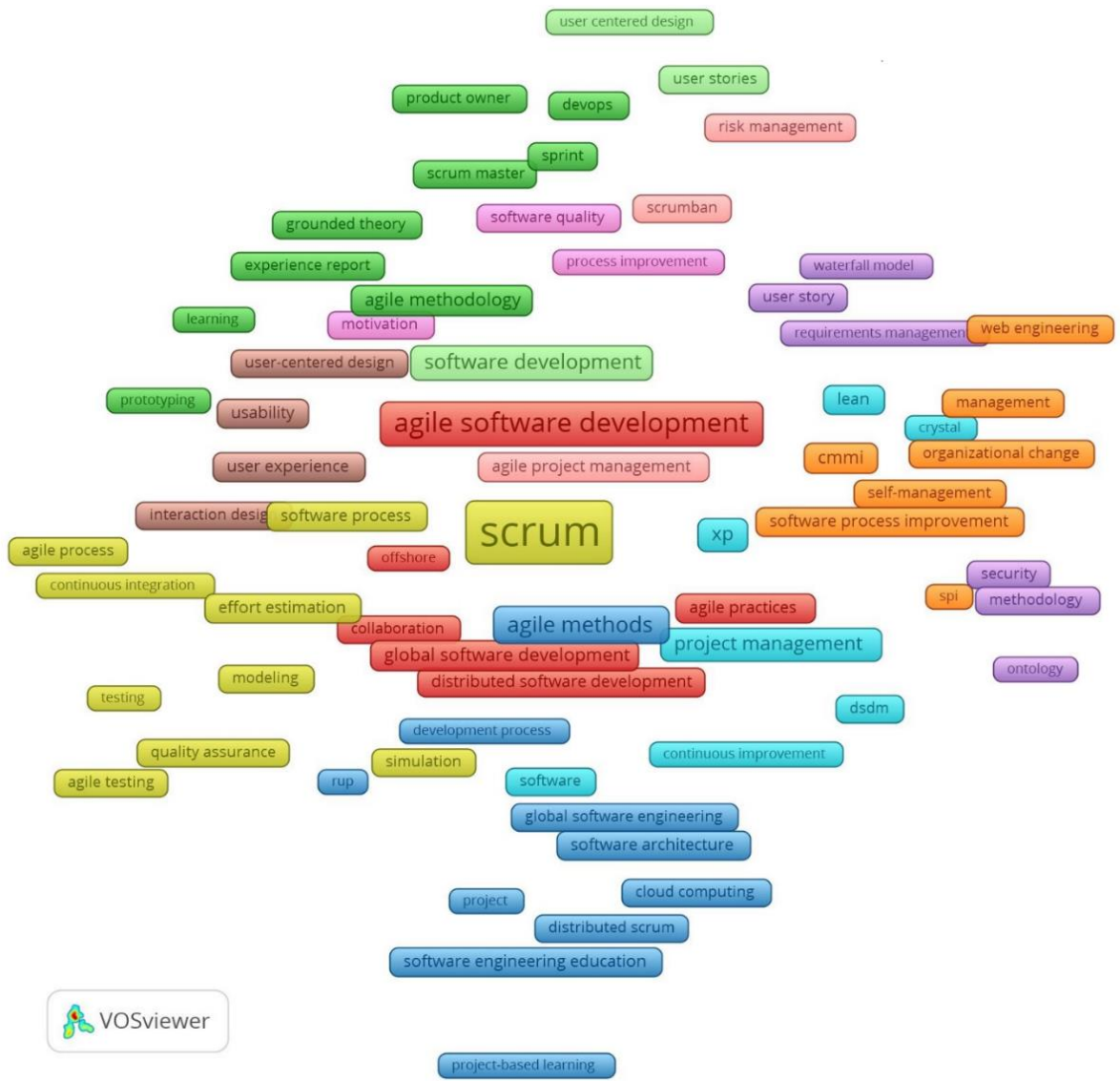

*Figure 3. The author keyword cluster landscape of the Scrum research*

*Table 1 Themes of Scrume research*

| Theme | No of publications | Landscape | Colour | Representative terms/author keywords used as codes in the thematic analysis |
|---|---|---|---|---|
| I. Improving software processes with education and training | 75 | Term | Violet | Teaching, education, training, skill, academia, Software process, software engineer, assessment, academia, learning, training, standard, iso iec |
| | | Keyword | Pink | Process improvement software quality, motivation |
| II. Interacting with users to develop working software | 144 | Term | Orange | User, persona, usability, user experience, interaction, agile manifesto, working software |
| | | Keyword | Brown | User centred design, interaction design, user experience, usability |
| III. Other software development methods and Scrum | 275 | Term | Yellow | Feature driven development; asd (agile software development), Extreme programming, software development methodology |
| | | Keyword | Light blue | Lean, crystal, XP, dsdm |
| IV. Translation of Scrum development aproach | 287 | Term | Green | Business value, stakeholder, software development company, continuous integration, enterprise, cmmi, software development lifecycle, mobile applications |
| | | Keyword | Orange | Organizational change, software process improvement, cmmi, self-management, web engineering |
| V. Interactive requirements engineering with user stories | 345 | Term | Light Blue | Agile software development team, customer requirement, user story, conflict, customer satisfaction, budget |
| | | Keyword | Green + violet | Agile methodology, software development, requirement management, user centered design, user story(ies), product owner, scrum muster, sprint, prototyping, learning, experience report, grounded theory, security, ontology |
| VI. Managing Scrum team in global software development | 105 | Term | Blue | Scrum team, product owner, scrum master, project manager, team member, coordination, global software development |
| | | Keyword | Red + Rose + Blue | Global software development, global software engineering, agile project management, agile software development, agile methods, distributed software development, distributed Scrum, risk management, collaboration, cloud computing, Scrumban, offshore, project based learning |
| VII. Scrum activities in every day work | 130 | Term | Red | Agile environment, scrum framework, infrastructure, business, software team, product backlog, client, daily meeting, code, trust, barrier, flexibility |
| | | Keyword | Yellow | Software process, agile process, scrum, effort estimation, modelling, simulation, agile testing, continuous integration |

The analysis of the most prolific papers for the methods, tools, roles, artefacts and objectives per theme resulted in 67 publications from the period 2007 to 2019. (Table 2). According to the publishing years, the most state of the art research is done in the themes related to process improvements and user developer interactions. The research seems to be the most established in the theme related to the use of Scrum in every day practice.

*Table 2a Methods. Tools, Roles, Objectives and Artefacts with most prolific publications*

| Theme | Methods | Tools | Roles | Objectives | Artefacts |
|---|---|---|---|---|---|
| I | Lean software development [48], Distributed Scrum [49] | Gamification [50], learning standards [51], Scrum framework [52] | Software engineer | Software process improvement [50], applicability [53], software quality [54] | software process [55], software model [56], standard [51] |
| II | | Agile manifesto [57], user centered design [57], interaction design [58] | User [57], Persona [59] | Usability [60] | Working software [58] |
| III | FDD [61], Crystal Clear [44], DSDM [45], Lean [62], Extreme programming [61], RUP [46], Kanban [1] | Effort estimation [63], global software engineering [64], | | Iterative development [65], uncertainty [66] | Software system |
| IV | Agile techniques [67] | Content analysis [68], knowledge management [69], guidance [70], CMMI [71], risk management [72], automation [47], cloud computing [73], self management [74], web engineering [75] | Project team [76], software developer [77] | Transparency [78], business value, functionality, dependency [79] | Software |
| V | Scrum framework [72] | User stories [80], failure prevention [81], user centered design [82], ontology [55], knowledge management [83] | Agile software development team [84] | Security [85], compliance [86], budget [87], customer satisfaction [88] | Software product |
| VI | SDLC [89], agile methodology [90] | Backlog [91], review [92], coordination [93], continuous improvement [94], sprint [95], prototyping [96], empirical studies [97], teamwork [98] | Scrum team [98], product owner [88], Scrum master [99], architect [100], project manager [101], team member [101] expert [102] | Responsibility [103], success factor [81], product quality [104], performance [95], agility [105] | Scrum project [101] |
| VII | Scrumban, COBIT [106], DevOps [107] | Pair programming [108], agile project management and development [109], Scrum framework [110], simulation, risk management [109], continuous integration [107], modelling [62], agile testing [111], quality assurance [112], effort estimation [113] | Software team [101], employee | Adaptability [89], flexibility [114], trust [115], efficiency [113] | |

# DISCUSSION

Our study showed that Scrum research is focusing on process improvement, user centred development, development of projects of any size (small to global), how to make development and management flexible and adaptive, roles of developers and users in the development process, how to translate Scrum into traditional software development organisations, how to educate Scrum users and how to use "untraditional methods" like grounded theory, games or content analysis to improve software processes. Those results are confirmed by review papers concerned with the agile development in general *[116]–[118]*. Contrary to our findings some author question the scalability of Scrum to large or very large projects *[119]–[121]*.

One of the more interesting aspects of the Scrum compared to the traditional approaches is the use of gaming in software development. Gaming is used during various activities, especially interesting is the application of gaming in the so called retrospectives – meetings devoted to process improvements [122]. During this activity games are used for data gathering, risk management, team cohesion, timeline assessment and priority of ongoing activities [123]. Another interesting approach to improve the Scrum processes not normally used in the traditional software engineering is the Grounded theory. It is used to support transition from a traditional software development to the Scrum based software development [124], and how an immature Scrum team can become self-organised [125], [126]. Another untraditional tool emerging in Scrum is the Content analysis which is used to better understand users [127].

While the traditional software engineering methods are mostly used as stand-alone, the Scrum is sometimes used in the combination with other development methods. Such hybrids include approaches where a traditional process model serves as (1) a framework in which several fine-grained Scrum practices are plugged in [128] or the traditional waterfall approach is incorporated into the Scrum to "manage chaos and risk" [129]. Other hybrids employ UML in user stories presentations [130] or use lean principles for continuous agile process improvement [131]. A recent more complex hybrid integrates Scrum, Kanban and Waterfall approaches aiming to support the automation of the software systems development [132]. Similar attempts have been made with Scrumban [133] and ScrumFall [134].

Recent publications also compared Scrum with other development approaches. An interesting comparison shows that Scrum, Kanban and Extreme programming requirements election is more effective comparing to other agile approaches, which often still adopted the traditional requirement process [135]. Another study showed that user experience is better if using Kanban than Scrum [136]. A recent study shows that synthesising Lean software development and distributed Scrum led to technical advances in system integration of network software systems [137]. Studies comparing Scrum and Waterfall revelled that using Scrum is preferable when issues should be fixed in a short period of time, accountability should be on very high level, when projects are not too large and when all the details are not known in the beginning of the projects, which all reults in higher Scrum execution success rates [138].

Few universities started including Scrum approach in their curricula [139], hence most of the Scrum education and training is done outside the regular university studies. Additionally, software engineering is a continuous process, which should incorporate needs of the software industry, thus virtual environments can serve as an appropriate training platform [140]. Other interesting training platforms are the gamification [141], [142], LEGO blocks [143], [144], doing real-life projects [145] and using learning platforms [54].

Our study also identified possible future directions in Scrum research. Collaborative games could be used first to foster the active user involvement in requirements engineering to envision new business and technical opportunities and shape the solutions [146] and second to revitalize the retrospective meetings [147]. Inter-team coordination employing a hybrid of agile and traditional planning tools could be used in large scale Scrum projects [148]. Similarly, knowledge management might be used in very large scale Scrum projects for knowledge transfer between teams [149] and to prevent knowledge loss [150]. The challenges of global software development [151] like self organization between remote teams [64] and bridging geographical, socio-cultural and temporal boundaries [152] will be tackled with the use of distributed software development [153] and model driven software architectures [154]. One of the future challenges is also how to trace non-functional requirement and the solution might be the integration of Feature Driven Development and Scrum [155]. User stories are a core approach in Scrum requirements engineering and it seems that future research will also be directed to them. How to estimate the needed resources based on the complexity of the user stories [156], conceptualisation of the consensus procedures to determine when a story is completed [80], reducing the ambiguity caused by using natural language in user stories [55] and deriving user for distributed Scrum [157] might be some of the challenges. Research on use of machine learning support to Scrum development will probably also receive more attention, for example in cost estimation [158], automatic testing [159] and general resources estimation [160].

**Challenges and limitations**

First, we used just one bibliographic database, namely Scopus the Elsevier's abstract and citation database of peer-reviewed literature. The reason was that Scopus is the largest of that kind of databases containing scientific journals, books and conference proceedings. It delivers a comprehensive overview of the world's research output in the fields of science including technology [161]. To maintain a certain degree of the quality of the corpus publication we didn't include the grey literature, which is mainly not peer-reviewed.

Second, to maintain broadness required for scoping we created a comprehensive database of existing literature on Scrum research, hence the volume of literature was to large to chart the publications in more depth. Consequently, we opted to produce a tabulated map of most prolific publications.

Finally, thematic analysis is a qualitative approach, thus susceptible to bias. While the authors done the study as objectively as possible, other authors might come to different results.

On the other hand our study is a first comprehensive bibliometric study focusing on solely on Scrum research. Additionally, the study was performed with a novel triangulating approach, combining bibliometric mapping and thematic analysis.

## CONCLUSION

The synthetic scoping review presents a holistic view on the Scrum research, as revealed by the bibliometric and bibliometric mapping based analysis of Scrum related publications indexed in the Scopus database. The review can help researchers and practitioners to understand the broader aspects of Scrum` research and its translation to practice. On the other hand, it can inform a novice, interested reader or software development professional without specific knowledge on Scrum to develop a perspective on the most important research themes, methods, tools and benefits of using Scrum. Our scoping review has also produced a comprehensive literature map, which points interested readers to most prolific publications.